\documentclass[prb,twocolumn,superscriptaddress,showpacs,amsmath,amssymb]{revtex4}
\usepackage{amsmath}
\usepackage{textcomp}
\usepackage{graphicx}
\usepackage{dcolumn}

\begin{document}
\title{Electronic and magnetic properties of superlattices of graphene/graphane nanoribbons with different edge hydrogenation}
\author{A. D. Hern\'andez-Nieves}
\email{alexande@cab.cnea.gov.ar}  \affiliation{Department of Physics, University of Antwerp,
Groenenborgerlaan 171, B-2020 Antwerpen, Belgium}
\affiliation{Centro Atomico Bariloche, 8400 San Carlos de Bariloche,
Rio Negro, Argentina }
\author{B. Partoens}
\affiliation{Department of Physics, University of Antwerp,
Groenenborgerlaan 171, B-2020 Antwerpen, Belgium}
\author{F.~M.~Peeters}
\email{francois.peeters@ua.ac.be} \affiliation{Department of
Physics, University of Antwerp, Groenenborgerlaan 171, B-2020
Antwerpen, Belgium}

\date{ \today }

\begin{abstract}
Zigzag graphene nanoribbons patterned on graphane are studied using spin-polarized ab initio calculations.
We found that the electronic and magnetic properties of the graphene/graphane superlattice strongly depends on the degree of 
hydrogenation at the interfaces between the two materials. When both zigzag interfaces are fully hydrogenated, the superlattice 
behaves like a freestanding zigzag graphene nanoribbon, and the magnetic ground state is antiferromagnetic. 
When one of the interfaces is half hydrogenated, the magnetic ground state becomes ferromagnetic, and the system is very close to being a 
half metal with possible spintronics applications whereas the magnetic ground state of the superlattice with both 
interfaces half hydrogenated 
is again antiferromagnetic. In this last case, both edges of the graphane nanoribbon also contribute to the total 
magnetization of the system. All the spin-polarized ground states are semiconducting, independent of the degree of
hydrogenation of the interfaces. The ab initio results are supplemented by a simple tight-binding analysis that captures 
the main qualitative features. Our ab initio results show that patterned hydrogenation of graphene is a promising way to 
obtain stable graphene nanoribbons with interesting technological applications.


\end{abstract}


\maketitle

\section{Introduction}

Graphene \cite{novoselov:2004} has emerged as a prospective material for nanoelectronics 
\cite{han:2007,li:2008,han:2010,jiao:2010} and spintronics applications,\cite{trombos:2007, karpan:2007} 
owing to the high mobility of 
carriers, the low-intrinsic spin-orbit interaction, as well as the low-hyperfine interaction of the electron spins with the 
carbon nuclei.\cite{trombos:2007}  The absence of an energy bandgap in graphene can be overcome by patterning graphene into 
nanometer size ribbons, which have been realized in recent experimental studies.\cite{han:2007,li:2008,han:2010,jiao:2010} 
The existence of width-dependent energy bandgaps makes graphene nanoribbons (GNRs) a potentially useful structure for various 
semiconducting applications. Ab initio calculations show that zigzag GNRs are 
magnetic\cite{kusakabe:2003,son:2006a} and they become half metallic by breaking the symmetry between the left and right 
edges.\cite{kusakabe:2003,kan:2008,son:2006b} This can be achieved by using a different chemical termination at each 
edge,\cite{kusakabe:2003,kan:2008} or by applying an external electric field.\cite{son:2006b}

Transport properties of GNRs depend on its edge disorder.\cite{han:2010} At present, controlling the shape and width of the 
GNR is a very challenging experimental task.\cite{han:2010,jiao:2010} 
There are also experimental evidences, corroborated by ab initio calculations, that the zigzag edge is metastable in vacuum 
because it shows a planar reconstruction to lower its energy.\cite{koskinen:2009} A new possibility for the fabrication of 
high-quality graphene nanoribbons is by selectively hydrogenating graphene or by carving GNRs on a graphane sheet.\cite{singh:2009}

First-principles calculations predicted that graphane is a wide-gap insulator. \cite{sofo:2007,lebegue:2009} 
This novel material was synthesized recently by exposing graphene to a cold hydrogen plasma.\cite{Elias:2009}
The promising properties of graphene/graphane structures have attracted recently a large interest.
\cite{singh:2009,li:2009, lu:2009, balog:2010}
It was shown that the energy bandgaps of both zigzag and armchair graphane nanoribbons increase as the nanoribbons become 
narrower.\cite{li:2009, lu:2009} Hybrid graphene/graphane nanoribbons in vacuum where also studied from ab initio calculations 
and it was shown that the energy bandgap is dominated by the graphene section of the nanoribbon.\cite{lu:2009} 
A bandgap opening in graphene, induced by the patterned absorption of atomic hydrogen onto the Moir\'{e} 
superlattice positions of graphene grown on a substrate, was recently found experimentally.\cite{balog:2010} Ab initio 
calculations 
indicate that the observed gap opening is due to the confinement effect in the residual bare graphene regions.\cite{balog:2010}

The edge states of graphene/graphane superlattices where recently considered in the tight-binding approximation. 
If the interface is oriented along a zigzag direction, edge states enhance effects 
related to the spin-orbit interaction.\cite{loss:2010} The stability of graphene/graphane nanoribbons was also studied from 
ab initio calculations, and it was predicted that the graphene/graphane nanoribbons are stable down to the limit of 
a single carbon chain. \cite{tozzini:2010}

Patterned fluoridation of graphene is another potential way for making graphene nanoribbons. Experimental evidences of graphene 
fluoridation was recently reported. \cite{cheng:2010, robinson:2010} Graphene fluoride is a two-dimensional material that
exhibits an insulating behavior similar to graphane, 
\cite{cheng:2010, robinson:2010} which can be also used to create superlattices of graphene nanoribbons.

In this paper, we carry out spin-polarized ab initio calculations to study the electronic and magnetic properties of 
superlattices of zigzag graphene/graphane nanoribbons (ZGGNRs). We study three cases: (i) when all the interfaces between the 
graphene/graphane nanoribbons are fully hydrogenated, (ii) when one edge of the nanoribbons is fully hydrogenated and the other 
edge is half-hydrogenated, and (iii) when all the interfaces of the superlattice are half hydrogenated (see Fig.~\ref{relaxed}). 
We found that the magnetic ground state of the ZGGNRs are semiconducting in the above three cases but the electronic and 
magnetic properties strongly depend on the degree of hydrogenation of the interfaces.

The calculations were performed with the QUANTUM ESPRESSO package \cite{QE} employing spin-polarized density-functional 
theory and the Perdew-Burke-Ernzerhoff exchange-correlation functional.\cite{pbe}
An ultra-soft pseudopotential description of the ion-electron interaction \cite{vanderbilt} was used together 
with a plane-wave basis set for the electronic wave functions and the charge density, with energy cutoffs of 40 Ry and 400 Ry,
respectively. The electronic Brillouin zone integration was sampled with a $16\times 1\times 1$ uniform {\it k}-point mesh and 
a Gaussian smearing of 0.01 Ry. The two-dimensional behavior of the graphene/graphane nanoribbons was simulated 
by adding a vacuum region of 12 \AA~ above the graphene/graphane sheets. All the structures were relaxed using a criteria 
of forces and stresses on atoms of 0.001eV/\AA~ and 0.5GPa, respectively.

\section{Results}

\begin{figure}
\includegraphics[width=\linewidth]{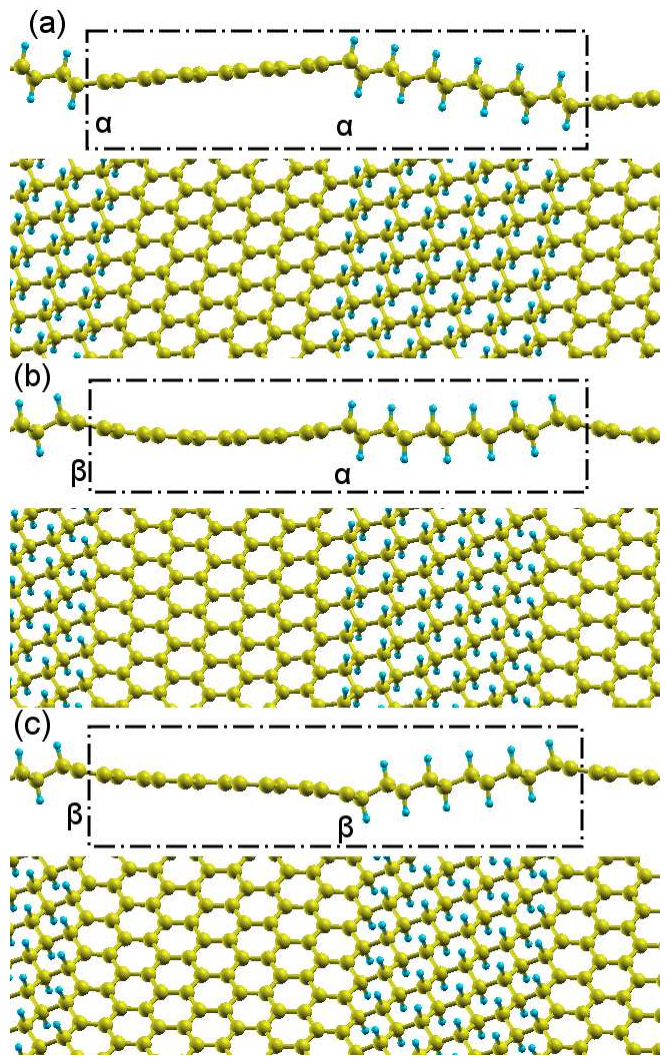}
\vspace{0cm} \caption{\label{relaxed}(Color online) Relaxed zigzag graphene/graphane nanoribbons. 
(a) Both graphene/graphane interfaces are fully hydrogenated (two $\alpha$ interfaces), (b) one of the interfaces is fully 
hydrogenated and the other one is half-hydrogenated ($\alpha\beta$ nanoribbon), and (c) both interfaces are 
half-hydrogenated resulting in $\beta\beta$ nanoribbons. Carbon atoms are represented by large (yellow) spheres and 
hydrogens atoms by small (light blue) spheres. We include a side view and a perspective view of the 
relaxed structures for each case. The unit cells of the superlattices are shown by dashed-dotted lines in the side views.}
\end{figure}

One or both of the two carbon atoms in each unit cell of the zigzag chain at the graphene/graphane interface can be hydrogenated 
(see Fig.~\ref{relaxed}). The degree of hydrogenation of the zigzag chains has a very important influence on the electronic 
properties of the ZGGNRs. In the following, we will refer to the fully covered hydrogen interface as an $\alpha$ interface 
and to the half-covered interface as a $\beta$ interface, as shown in Fig.~\ref{relaxed}. 
In principle, if we allow the hydrogen deposition to start on a certain sublattice of graphene, it would be possible to 
experimentally control the degree of hydrogenation of the interfaces.

Zigzag graphene nanoribbons (ZGNR) are usually classified by the number of zigzag chains across the ribbon 
width.\cite{son:2006a,son:2006b} In order to simplify the analysis, we considered superlattices of zigzag 
graphene/graphane nanoribbons where both, graphene and graphane nanoribbons, are six-chains wide. 
This means that the unit cell of Fig. ~\ref{relaxed}(a) has 24 carbon and 12 hydrogen atoms. 
The degree of hydrogenation at the interface, and therefore the number of hydrogen atoms, decreases from Fig.~\ref{relaxed}(a)\--\ref{relaxed}(c).

The interface between graphene and graphane nanoribbons is an interface between {\it sp$^2$} and {\it sp$^3$} bonded carbon atoms. 
Carbon atoms in graphene have a {\it sp$^2$} hybridization and the structure stabilizes as a flat sheet. 
The hybridization changes from {\it sp$^2$} to {\it sp$^3$} in graphane due to the presence of alternating hydrogen atoms placed on both 
sides of the carbon plane. The {\it sp$^3$} hybridization also forces nearest-neighbor carbon atoms in graphane to lie in different 
planes, and the resulting graphane structure is undulating.\cite{sofo:2007}

Relaxed structures of the ZGGNRs are shown in Fig.~\ref{relaxed}. All structures were fully relaxed according 
to the forces and stresses on the atoms by using spin-polarized calculations. 
When both interfaces are fully hydrogenated (two $\alpha$ interfaces) one side of the graphene nanoribbon 
is pushed up and the other side is pushed down by the graphane 
{\it sp$^3$}-bonded carbon atoms. This induces a tilt angle of 164\textdegree\; between adjacent graphene/graphane nanoribbons as shown 
in Fig.~\ref{relaxed}(a). To satisfy the conditions imposed by the {\it sp$^3$}/{\it sp$^2$} carbon interface, the graphene nanoribbon is tilted 
with respect to the graphane plane but it remains flat. A similar situation occurs when both interfaces are half-hydrogenated 
(two $\beta$-type interfaces) as shown in Fig.~\ref{relaxed}(c). In this case, the tilt angle between the graphene and 
graphane nanoribbons is 166.6 degrees. When one interface is fully hydrogenated and the other one is half-hydrogenated 
($\alpha\beta$ superlattice), the graphene nanoribbon is pushed down at both edges by the graphane {\it sp$^3$}-bonded carbon atoms. 
This results in a curvature of the graphene nanoribbons as shown in Fig.~\ref{relaxed}(b).

It is known that a freestanding $\alpha\beta$ graphene nanoribbon is not stable. We found that by using spatially selective 
hydrogenation of graphene we are able to create stable $\alpha\beta$ nanoribbons. The stability of those nanoribbons was checked 
by allowing relaxation of the structure in our ab initio calculations.

We also addressed the question about the feasibility of appearance of the zigzag interface by comparing it with the armchair 
interface. We prepared two supercells (one with a zigzag and the other one with an armchair arrangements of atoms) forming 
graphene/graphane nanoribbons and containing the same number of atoms (108 atoms), 72C and 36H. The zigzag supercell 
is equivalent to three times the unit cell of Fig.~\ref{relaxed}(a) replicated along the graphene/graphane nanoribbon. 
We obtained a very small energy difference between these zigzag and armchair graphene/graphane 
nanoribbons, only 0.002 eV/atom. This value is smaller than our numerical accuracy, and it is also very small if we compare it 
with the thermal energy at room temperature, 0.025eV. We can conclude that the formation energy of zigzag and 
armchair graphene/graphane nanoribbons are equally favorable at room temperature. This situation is different from the case of  
freestanding nanoribbons where it was found that the zigzag edge is metastable, because it shows a planar reconstruction 
to lower its energy.\cite{koskinen:2009} This shows that graphene/graphane superlattices are a more convenient way for 
stabilizing zigzag graphene nanoribbons.

In calculations not shown here, we carried out structure relaxations where the volume of the unit cell is kept constant 
and equal to the volume of an equivalent graphane system and where only the forces between atoms are relaxed. 
This is equivalent to the experimental situation where graphene nanoribbons are carved in a graphane sheet with fixed borders. 
In this case, we found no tilt angles between the graphene and graphane nanoribbons and only a very small curvature for the 
$\alpha\beta$ superlattice. The electronic or magnetic properties of the system did not change in a significant way if we compare 
with the fully relaxed case.

\begin{figure}[htp]
\includegraphics[width=\linewidth]{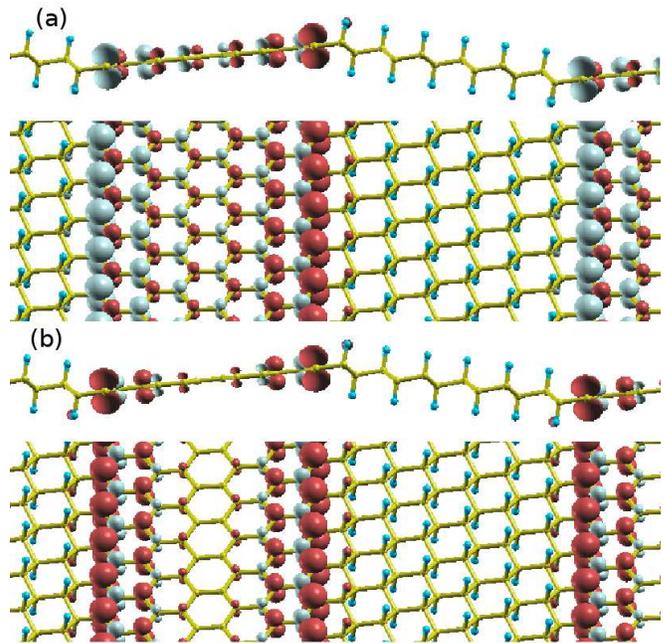}
\vspace{0cm} \caption{\label{SP_AA}(Color online) Three-dimensional (3D) graph of the spin density ($\rho_{\uparrow} - \rho_{\downarrow}$) 
of the $\alpha\alpha$ zigzag graphene/graphane nanoribbon. (a) Spin density of the antiferromagnetic ground state and 
(b) spin density of the metastable ferromagnetic state. In both cases, the magnetization is almost completely localized 
in the graphene nanoribbons, and the edges of the graphane nanoribbons are only slightly magnetized.
Regions with positive spin density are represented in red (dark gray) and regions with a negative spin density are shown in light blue (light gray).
Isosurfaces corresponds to ($\rho_{\uparrow} - \rho_{\downarrow}$)=0.02 $e/\AA^3$, where {\it e} is the electron charge.}
\end{figure}

In the following sections, we discuss separately the electronic and magnetic properties of the ZGGNRs with different degree 
of hydrogenation at the interfaces.

\subsection{Two fully hydrogenated interfaces: $\alpha\alpha$ zigzag graphene/graphane nanoribbons}

We first analyze the case where both zigzag interfaces are fully hydrogenated. The ground state of the system corresponds to 
an antiferromagnetic (AFM) ordering between the two opposite graphene edges and a ferromagnetic (FM) ordering along the zigzag 
carbon chains. We will refer to this state as an AFM state. 
The spin density ($\rho_{\uparrow} - \rho_{\downarrow}$) of the AFM state is represented in 
Fig.~\ref{SP_AA}(a) while the metastable FM state (ferromagnetic ordering both between the two graphene edges and along the 
zigzag carbon chains) is represented in Fig.~\ref{SP_AA}(b).

\begin{figure}[htp]
\includegraphics[width=\linewidth]{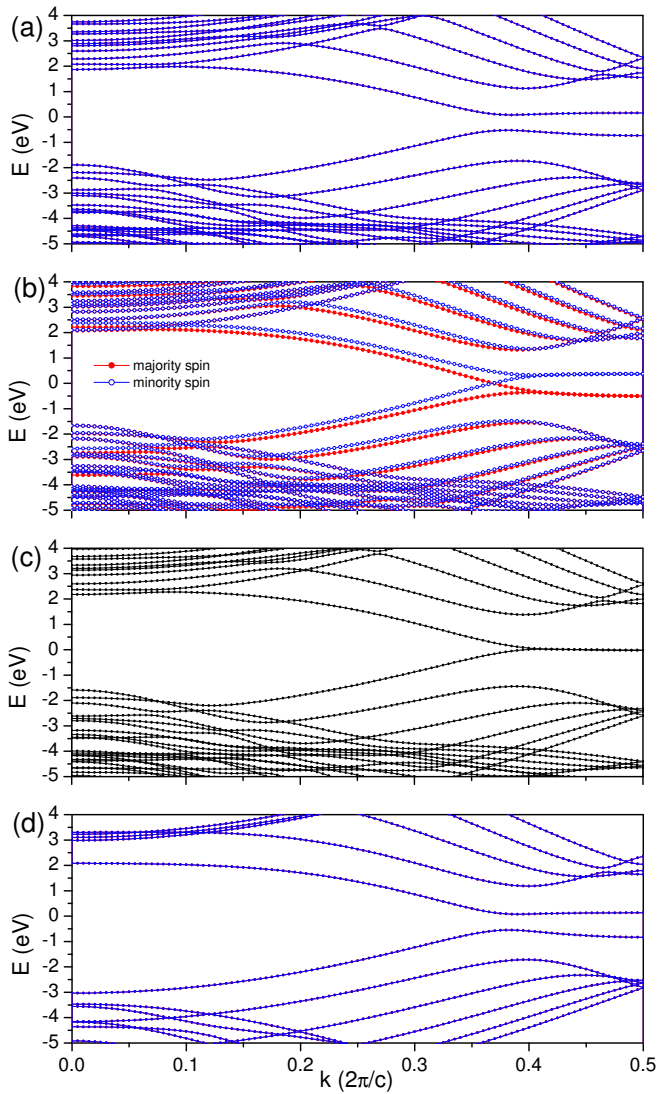}
\vspace{0cm} \caption{\label{bands_AA}(Color online) The spin-resolved band structures of the superlattice with all the interfaces fully 
hydrogenated, $\alpha\alpha$ ZGGNR superlattice. 
(a) Antiferromagnetic semiconducting ground state, (b) ferromagnetic metallic metastable state, 
and (c) nonmagnetic metallic metastable state. (d) The band structure of the AFM state of a hydrogen-passivated six-chains wide 
zigzag graphene nanoribbon. In all figures, the Fermi energy is set to zero.
The energy differences of the ferromagnetic (b) and the nonmagnetic states (c) with respect to the antiferromagnetic state (a) 
are $E_{AFM}-E_{FM}=8$meV/edge atom and $E_{AFM}-E_{NM}=35$meV/edge atom, respectively. Spin up and spin down bands coincide in (a), (c), and (d).}
\end{figure}

The total magnetic moment and the absolute magnetic moment for the 
FM state are 0.59$\mu_{B}$/cell and 0.95$\mu_{B}$/cell, respectively ($\mu_{B}$ is the Bohr magneton). The AFM state has an absolute magnetic moment of 1.09$\mu_{B}$/cell. 

The spin density ($\rho_{\uparrow} - \rho_{\downarrow}$) of the $\alpha\alpha$ ZGGNR are 
mostly concentrated in the graphene nanoribbons as shown in Figs.~\ref{SP_AA}(a) and  \ref{SP_AA}(b). 
When moving away from the interface to the center of the graphene nanoribbons, the spin density decreases 
faster in the FM state than in the AFM state. As a consequence, in the middle of the graphene nanoribbon, 
the spin density is lower in the FM state than in the AFM state. This explains the difference of 0.14$\mu_{B}$/cell 
in the absolute magnetization between both magnetic states. 

\begin{table}[htp]
\caption{Comparison between a six-chains wide freestanding zigzag graphene
nanoribbon (ZGNR) and a six-chains wide 
$\alpha \alpha$ zigzag graphene/graphane nanoribbon ($\alpha \alpha$ ZGGNR).
Energies are in millielectron volt per edge atom, bandgaps in millielectron volt, and the absolute magnetic
moments ($\mu_{abs}$) in $\mu_{B}$/cell. The total magnetic moments for the FM
state are 0.51$\mu_{B}$/cell and 0.59$\mu_{B}$/cell for ZGNR and
$\alpha\alpha$ZGGNR, respectively.\\}

\begin{tabular}{c|ccc|ccc}
\hline \hline
    &$\alpha \alpha$ ZGGNR & & &ZGNR&\\
\hline
\;state\;    &Bandgap    &$\;\Delta E\;$    &$\;\;\mu_{abs}\;\;$  & \;Bandgap\; &$\;\Delta E\;$        &$\;\;\mu_{abs}\;\;$\\
\hline
AFM &578    & - &1.09 & 607 & -   & 1.06 \\
FM  &0.0    &8  &0.95 &  0.0  & 9   &0.84\\
NM  &0.0    &35 &0.0  &  0.0  & 36  &0.0 \\

\hline
\hline
\end{tabular}
\label{table1}
\end{table}

We also carried out ab initio calculations with a unit cell twice larger along the zigzag carbon chains to explore whether an 
antiferromagnetic ordering along the nanoribbons can be the ground state of the system. However, we found that ferromagnetic 
order along the nanoribbon shown in Figs.~\ref{SP_AA}(a) and  \ref{SP_AA}(b) gives always lower energy states.

The band structure of the ZGGNRs with the two interfaces fully hydrogenated are shown in Fig.~\ref{bands_AA}. 
Figure~\ref{bands_AA}(a) corresponds to the semiconducting AFM ground state.  The FM metastable state [Fig.~\ref{bands_AA}(b)] 
shows a metallic behavior. As the graphane nanoribbons are not spin polarized there is no significant spin density 
observed that can be associated with the graphane regions. The nonmagnetic metastable state is a semimetal as shown in 
Fig.~\ref{bands_AA}(c). 

\begin{figure} [ht]
\includegraphics[width=\linewidth]{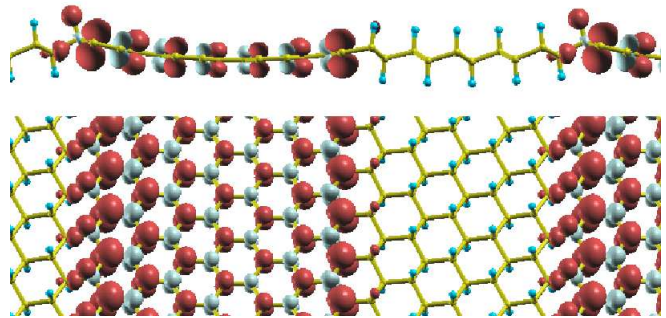}
\vspace{0cm} \caption{\label{SP_AB}(Color online) Side and perspective views of spin density of 
the ferromagnetic ground state of the $\alpha\beta$ ZGGNRs. Around the $\beta$ interface the spin density is larger than 
in the $\alpha$ interface. 
The $\beta$-edge of the graphane nanoribbon is also magnetized, while the spin-state of the $\alpha$ edge is similar to 
Fig.~\ref{SP_AA}. Regions with positive spin density are represented in red (dark gray) and regions with a negative spin density 
are shown in light blue (light gray). Isosurfaces corresponds to ($\rho_{\uparrow} - \rho_{\downarrow}$)=0.02 $e/\AA^3$.}
\end{figure}

The band structure of the AFM state of an equivalent six-chains wide ZGNR, thus without graphane, is represented in Fig.~\ref{bands_AA}(d). 
We see that, near the Fermi level, the band structure of the AFM state of the superlattice [Fig.~\ref{bands_AA} (a)] is very 
similar to the band structure of the AFM state of a graphene nanoribbon in vacuum [Fig.~\ref{bands_AA}(d)].
This similarity is also observed in all the other metastable states. Thus, near the Fermi level, the band structure of the 
superlattice is dominated by the contribution of the graphene nanoribbons. 

We carefully compared the magnetic behavior of isolated six-chains wide ZGGNR with graphene nanoribbons 
embedded in graphane. The comparison between their electronic and magnetic properties are summarized in Table \ref{table1}. 
We found very small differences in the energy bandgaps, magnetic moments, and energy differences between the different 
magnetic states of graphene nanoribbons in vacuum and patterned in graphane. 
These results show that patterning graphene nanoribbons in graphane is a promising way of obtaining zigzag 
graphene nanoribbons with the same electronic and magnetic properties as in vacuum.

In the four cases represented in Fig.~\ref{bands_AA}, localized edge states generate flat bands near the Fermi level. 
Information about the spatial localization of the edge states can be obtained by analyzing the partial (or band decomposed) 
charge density (PCD). This magnitude is also called integrated local density of states (ILDOS), which describes the density of 
states but space resolved, and it is useful when interpreting the data from STM experiments. In the following, we will use the 
term PCD. 

\begin{figure} [ht]
\includegraphics[width=\linewidth]{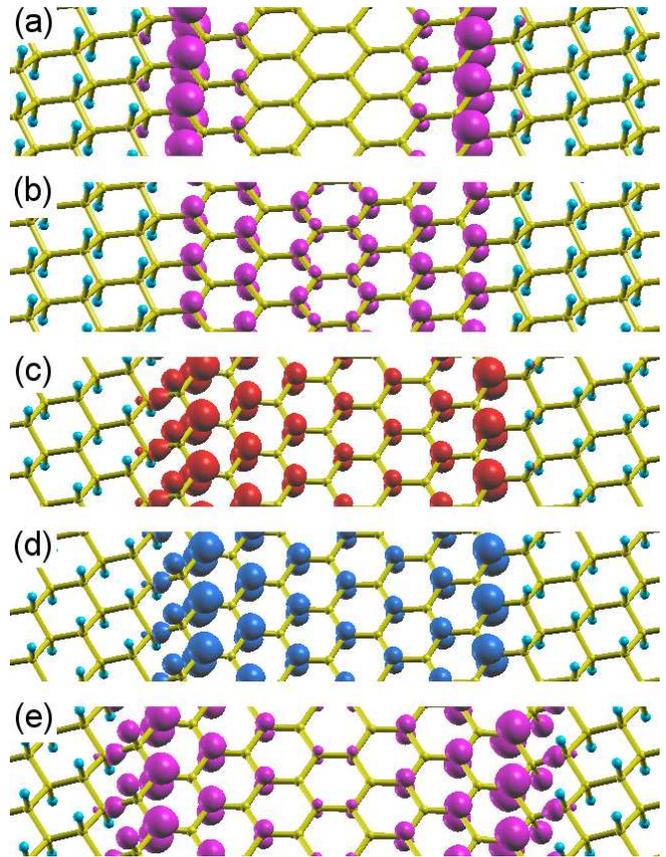}
\vspace{0cm} \caption{\label{PCD}(Color online) 3D graph of the PCD. 
(a) PCD of the nonmagnetic state of an $\alpha \alpha$ superlattice integrated in the energy window $E_f \pm 0.02$eV, 
(b) the same as before but from $E_f-0.09$eV to $E_f-1$eV, (c) PCD of the majority spin states of an 
$\alpha \beta$ superlattice in the FM state integrated in $E_f \pm 0.7$eV, (d) the same as (c) but for the 
minority spin states, and (e) PCD of the AFM state of a $\beta \beta$ superlattice from $E_f$ to $E_f-0.16$eV. 
Red regions corresponds to majority spins (c), blue to minority spin (d), and violet means that both spins components were 
included in (a), (b), and (e).}
\end{figure}

Figure~\ref{PCD}(a) shows the spatial localization or PCD of the edge states 
around $E_f$ ($E_f\pm 0.02$eV) for the 
nonmagnetic case. If we exclude the states at $E_f$, and do an integration from $E_f-0.09$eV to $E_f-1$eV, 
we see that the states become less localized, see Fig.~\ref{PCD}(b). 
An interesting feature in Fig.~\ref{PCD}(b) is that near each edge of the graphene nanoribbon only one of the two carbon 
sublattices makes a large contribution to the states near the Fermi level. 
In fact, we can appreciate from Fig.~\ref{PCD}(b) that different carbon sublattices contribute at both edges while just in the 
middle of the graphene nanoribbon both sublattices contribute to the bands. The main contribution of the other carbon sublattice 
is located near the center of the Brillouin zone at the $\Gamma$ point. These states lie at lower energies ($E_f-2$eV) 
and are entangled with the bands coming from the graphane nanoribbon.

All the above mentioned features of the band structure and the PCD of $\alpha\alpha$ ZGGNRs can be captured  qualitatively with 
a simple tight-binding model, which results are summarized in Figs.~\ref{Graph9}(a) and \ref{Graph9}(d) of the Appendix.

\subsection{Superlattice with one interface fully hydrogenated and the other one half-hydrogenated: $\alpha\beta$ ZGGNRs}

When one of the interfaces is half-hydrogenated and the other one is fully hydrogenated the spin-polarized ground state of the 
system corresponds to a ferromagnetic ordering, both along the graphene/graphane interface and between the two edges of the 
graphene nanoribbon.  
Figure~\ref{SP_AB} shows the spin density for the FM ground state of the $\alpha \beta$ ZGGNR. 

If we compare with an $\alpha \alpha$ ZGGNR, the absolute magnetization increases to 1.73$\mu_{B}$/cell and the total magnetization to  
1.0$\mu_{B}$/cell in the $\alpha \beta$ ZGGNR. The side view of the spin density in Fig.~\ref{SP_AB} shows that 
the $\beta$-edge of the graphane nanoribbons is also spin polarized.

\begin{figure} [htp]
\includegraphics[width=\linewidth]{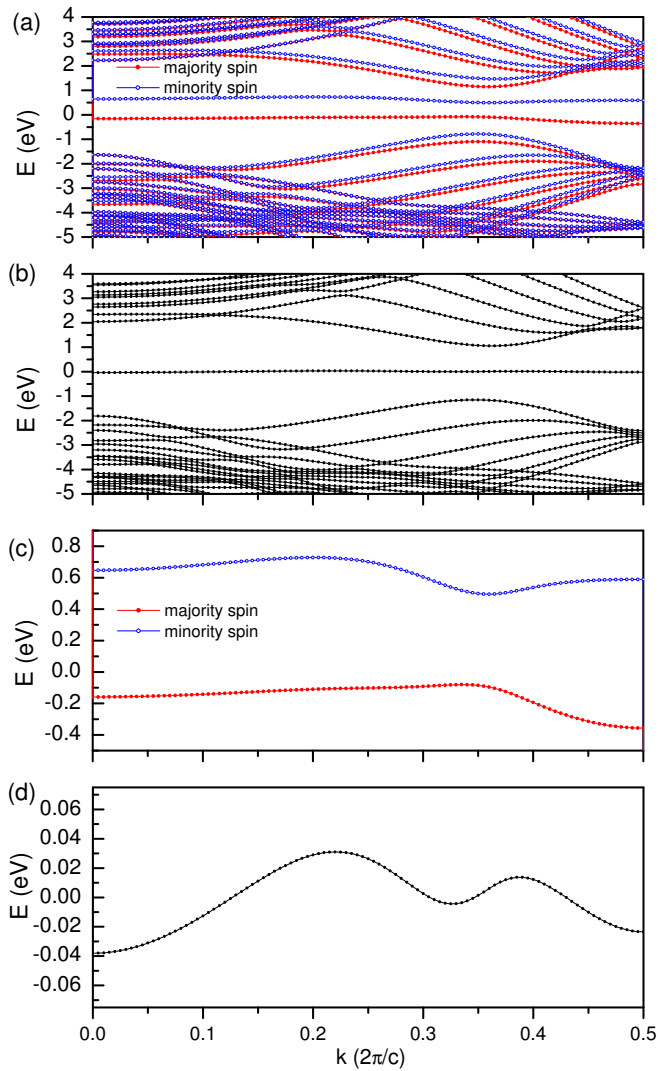}
\vspace{0cm} \caption{\label{bands_AB}(Color online) The spin-resolved band structures of the $\alpha\beta$ zigzag graphene/graphane nanoribbons. 
(a) ferromagnetic semiconducting ground state and (b) nonmagnetic metallic metastable state. (c) and (d) are zooms of (a) and 
(b) near the Fermi level, which was set to zero.
The energy difference between the ferromagnetic and the nonmagnetic state is $E_{FM}-E_{NM}=76$meV/edge atom. }
\end{figure}

The band structure of the $\alpha\beta$ ZGGNRs are shown in Fig.~\ref{bands_AB}. 
The FM ground state shows a semiconducting behavior [Fig.~\ref{bands_AB}(a)]. There are two spin-polarized flat 
bands near the Fermi level with a direct bandgap of 0.58eV. Ferromagnetic semiconductors have attracted a great interest 
since 1990s because 
they may lead to  spintronics applications.\cite{jungwirth} $\alpha\beta$ zigzag graphene/graphane nanoribbons may be an 
alternative for the well-studied diluted magnetic semiconductors (DMS) materials \cite{jungwirth} with the fundamental 
difference that the system is purely two-dimensional in this case. The band structure corresponding to the nonmagnetic 
metastable state is represented in Fig.~\ref{bands_AB}(b). 

If we compare the energy difference between the magnetic and the nonmagnetic cases for the three different types of nanoribbons 
shown in Table~\ref{table2}, we can see that the largest energy difference between the nonmagnetic and magnetic cases occurs 
for $\alpha\beta$ ZGGNRs. This indicates that the FM state of $\alpha\beta$ ZGGNRs could be even more stable than the other 
cases considered in this work and summarized in Table~\ref{table2}.

\begin{table}[htp]
\caption{Electronic and magnetic properties of graphene/graphane nanoribbons
with different hydrogenation at the interfaces.
Bandgaps are in millielectron volt, energy differences ($\Delta E$) are in millielectron volt per edge atom (meV/e.a.), 
and the total and absolute
magnetic moments ($\mu_{t}$ and $\mu_{abs}$, respectively) are in units of $\mu_{B}$/cell.\\}

\begin{tabular}{c|ccccc}
\hline \hline
ZGGNR & state    &\;\;Bandgap\;\;&\;\;$\Delta E$\;\;&$\mu_{t}$       &\;\;$\mu_{abs}$ \;\;\\ 
      &          &   (meV)       & (meV/e.a.)  &($\mu_{B}$/cell)&  ($\mu_{B}$/cell)\\ 
\hline
$\alpha\alpha$ &  AFM & 578    & 0   & 0.0    & 1.09   \\ 
               &  FM  & 0.0    & 8   & 0.59   & 0.95  \\ 
               &  NM  & 0.0    & 35  & 0.0    &  0.0  \\ 
\hline
$\alpha\beta$  &  FM & 580    & 0   & 1.0    & 1.73  \\ 
               &  NM & 0.0    & 76  & 0.0    &  0.0  \\ 
               & AFM &  -     & \footnote{We could not stabilize the AFM state for $\alpha\beta$ ZGGNRs. This could indicate that this state is not metaestable. }   
                                    &  -     & -   \\ 
\hline
$\beta\beta$   & AFM & 760    & 0   & 0.0   & 1.92   \\ 
               &  NM & 290    & 37  & 0.0   &  0.0  \\ 
               &  FM &  -     & \footnote{The same situation ocurred for the FM state of $\beta\beta$ ZGGNRs.}   
                                    &  -    &  - \\ 
\hline
\hline
\end{tabular} 
\label{table2}
\end{table}

$\alpha\beta$ ZGGNRs have an extra carbon atom and, therefore, an odd number of atoms in the graphene nanoribbons 
(13 carbon atoms in this case). This adds up one extra band near the Fermi level, which can be appreciated by comparing the 
nonmagnetic states of 
Figs.~\ref{bands_AA}(c) and \ref{bands_AB}(b). The two bands that were previously dominating the behavior near the Fermi level 
in $\alpha\alpha$ ZGGNRs are now, respectively, pushed up and down in energy and very flat bands appear near the Fermi level, 
both for the 
FM ground state of Fig.~\ref{bands_AB}(a) and for the nonmagnetic metastable state of Fig.~\ref{bands_AB}(b).

Figures~\ref{bands_AB}(c) and ~\ref{bands_AB}(d) are zooms of the states near the Fermi level for the FM and non magnetic 
states, respectively. Even when the bands are very flat, there is a small dispersion that is caused by the 
interaction with the graphene/graphane interfaces. 

The results of the simple tight-binding model shown in Figs.~\ref{Graph9}(b) and \ref{Graph9}(e) 
reproduce qualitatively the behavior of the nonmagnetic band structure. However, the band at the Fermi level is completely flat. 
In order to obtain dispersion in this band, we need to include an interaction with the graphane nanoribbon. This can be done by 
introducing an extra row of carbon atoms coupled with hydrogen as in graphane, which simulates the 
graphene/graphane interface. In doing so we find an extra band with a dispersion comparable to those from our ab initio 
simulations. The band width does not appreciably depend on the width of the graphene nanoribbon.

\begin{figure} [ht]
\includegraphics[width=\linewidth]{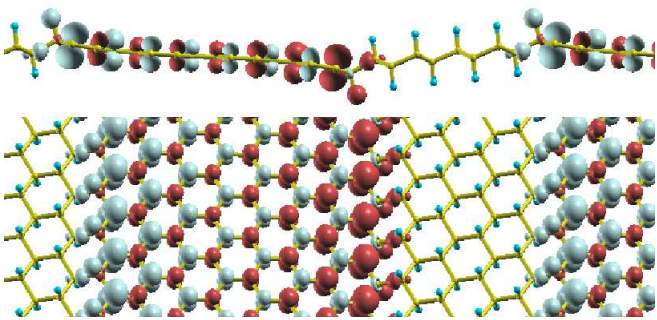}
\vspace{0cm} \caption{\label{SP_BB}(Color online) Side and perspective views of the 3D graph of the spin density of the 
antiferromagnetic ground state of  the $\beta\beta$ graphene/graphane nanoribbons. The spin-polarized regions extend to include both edges of the graphane nanoribbon. 
Regions with positive spin density are represented in red (dark gray) and regions with a negative spin density are shown in 
light blue (light gray). Isosurfaces corresponds to ($\rho_{\uparrow} - \rho_{\downarrow}$)=0.02 $e/\AA^3$.}
\end{figure}

More complicated tight-binding models for the case of an $\alpha\beta$ ZGGNRs where introduced in 
Ref.~\onlinecite{loss:2010}. However, the obtained band width  \cite{loss:2010} was four times larger 
than the one found with our ab initio results.

A very interesting feature of the PCD of the FM ground state of Figs.~\ref{PCD}(c) and \ref{PCD}(d) is that only one of 
the carbon sublattices is contributing to the edge state. The same happens for the nonmagnetic metastable state (not shown here). 
This can be incorporated in a simple tight-binding model represented in Fig.~\ref{Graph9}(e). 
All the coefficients corresponding to one of the carbon sublattices are zero, B=D=F=H=J=L=0 in Fig.~\ref{Graph9}(e).

Notice the nearly parabolic feature of the band of the nonmagnetic state of Fig.~\ref{bands_AB}(d) 
at {\it k}=0 and {\it k}=0.5. This allows to define and effective mass ($m_{eff}$) for the electrons, 
in this case $m_{eff}=5.77m_e$ for {\it k}=0 and $m_{eff}=3.22 m_e$ for {\it k}=0.5, where $m_e$ is electron rest mass.

\begin{figure} [ht]
\includegraphics[width=\linewidth]{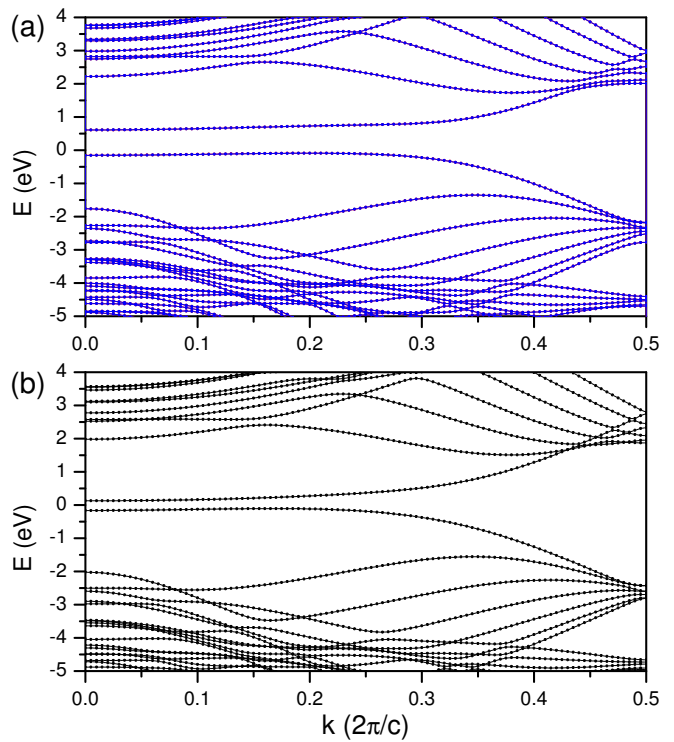}
\vspace{0cm} \caption{\label{bands_BB}(Color online) The spin-resolved band
structures of a $\beta\beta$ ZGGNRs. (a) Antiferromagnetic semiconducting ground
state and (b) nonmagnetic semiconducting metastable state. The Fermi energy is
set to zero. The energy difference between the antiferromagnetic and the
nonmagnetic state is $E_{AFM}-E_{NM}=37$meV/edge atom. 
The spin-polarized bands coincide in both cases.}
\end{figure}

\subsection{Superlattice with the two edges half-hydrogenated: $\beta\beta$ zigzag graphene/graphane nanoribbons}

We consider now the case of the $\beta\beta$ superlattice represented in Fig.~\ref{relaxed}(c).
The spin-polarized ground state is the AFM state as was also the case in the $\alpha\alpha$ superlattice.
Figure~\ref{SP_BB} shows the spin density for the AFM ground state of the $\beta \beta$ ZGGNR. Both edges of the graphane 
nanoribbon contributes to the spin polarization of the system. This can be better appreciated in the side view at the top of 
Fig.~\ref{SP_BB}. The absolute magnetic moment of the AFM state is 1.92 $\mu_{B}$/cell, almost twice larger than in the AFM 
state of the $\alpha\alpha$ ZGGNR (1.09 $\mu_{B}$/cell).

Figure~\ref{bands_BB}(a) shows the band structure of the AFM ground state of the $\beta\beta$ ZGGNR and the nonmagnetic 
metastable state is shown in Fig.~\ref{bands_BB}(b). Both are semiconducting states while the energy band gap of the AFM 
state (0.76eV) is larger than in the nonmagnetic case (0.29eV). It is interesting to note that even when we have a nonmagnetic 
state with a bandgap there is a transition to a magnetic state by further increasing the bandgap. The properties of both states 
are summarized in Table~\ref{table2}.

The partial charge density graph of the AFM phase near the Fermi level obtained from $E_f$ to $E_f-0.16$eV shows that the edge states of 
the $\beta\beta$ superlattice extend also to the edge of the graphane nanoribbon as shown in Fig.~\ref{PCD}(e). The edge 
states decay more rapidly in the graphane nanoribbon than in the graphene region when we move away from the interface. 

The qualitative features of the $\beta\beta$ ZGGNRs are also captured by the tight-binding model of the 
Appendix as can be appreciated from Figs.~\ref{Graph9}(c) and \ref{Graph9}(f). Figure~\ref{Graph9}(f) also correlates with the 
ab initio PCD of Fig.~\ref{PCD}(e). This is discussed in more detail in the Appendix. We see that the different carbon sublattices contribute differently near each edge. 
This situation is  similar to the case we explained before for the $\alpha\alpha$ ZGGNRs.

\section{Conclusions}

The degree of hydrogenation of the graphene/graphane interface has an important influence on the electronic and magnetic 
properties of the zigzag graphene/graphane nanoribbons. This parameter could be controlled experimentally to tune the bandgap 
and the magnetic properties of graphene/graphane superlattices.

We found that independent of the degree of hydrogenation of both interfaces the ground state of the system is semiconducting. 
However, the band structure near the Fermi level changes dramatically due to the contribution of the edge states in the three 
different superlattices. While the ground state of the system is antiferromagnetic in $\alpha\alpha$ and $\beta\beta$ ZGGNRs, 
it is ferromagnetic for $\alpha\beta$ ZGGNRs.

Our results show that $\alpha\alpha$ ZGGNRs are a promising way to obtain stable graphene nanoribbons with similar electronic and 
magnetic properties as freestanding zigzag nanoribbons. The basic properties of freestanding zigzag graphene 
edges are preserved when the system binds on both edges to insulating graphane regions.
We also found that the energy difference between zigzag and an armchair graphene/graphane nanoribbons is very small, the energy 
difference per atom is 
one order of magnitude smaller than the thermal energy at room temperature. This means that the formation energies of zigzag and 
armchair graphene/graphane nanoribbons are equally favorable at room temperature. This situation is different from the case 
of freestanding nanoribbons where it was found that the zigzag edge 
is metastable, because it shows a planar reconstruction to lower its energy.\cite{koskinen:2009} Then, graphene/graphane 
superlattices seem a promising way to solve structural and stability problems faced by freestanding nanoribbons.

We proposed a single atom thick ferromagnetic semiconductor in $\alpha\beta$ zigzag graphene/graphane nanoribbons.
It may be an alternative for the well-studied DMS materials \cite{jungwirth} with the fundamental difference that the system is purely two-dimensional in this case.
Diluted semiconductors are of interest from both fundamental and applied points of view, and may soon lead to spintronics devices with new functionalities. \cite{jungwirth}

$\alpha\beta$ ZGGNRs can also be analyzed in a more general framework. In order to render a ZGNR half metallic, 
the symmetry between the left and right edges of the zigzag nanoribbon should be broken. For example, in Ref. 
\onlinecite{son:2006b} this was done by applying an electric field to the system, in Ref. \onlinecite{kusakabe:2003} this was 
done by adding a Klein's edge to one side of the nanoribbon, and in Ref. \onlinecite{kan:2008} the two zigzag edges were terminated with different 
terminal groups, such as NO$_2$ and CH$_3$. The present work shows that in graphene/graphane superlattices, this could be 
achieved by controling the degree of hydrogenation at the two edges.

From Fig.~\ref{bands_AB}(c) we can appreciate that the ferromagnetic ground state of $\alpha\beta$ ZGGNRs is almost a 
half metal. As the Fermi level is in between two spin-polarized bands, with little charge doping the 
Fermi level can be shifted either to the spin-up or spin-down band. This could be used in a potentially interesting 
electronic device. So half-metalicity in $\alpha\beta$ ZGGNRs can be achieved not with the help of an external electric 
field as in Ref. \onlinecite{son:2006b} but with charge doping, which could offer new technological alternatives.

It is known that the width of the nanoribbons has a strong influence on the bandgap of the graphene nanoribbons 
\cite{son:2006a,son:2006b} as well as on the bandgap of the graphane nanoribbons.\cite{li:2009, lu:2009}
Both, the width of the graphene nanoribbons and the level of hydrogenation at the interface can be used to control 
the electronic and magnetic properties of graphene/graphane nanoribbons. While the width of the graphane region 
can be used to control the energy window where graphene and graphane bands entangle.

Most of the qualitative features obtained in our ab initio calculations can be explained with a simple tight-binding model 
incorporated in the appendix. This shows that the present results can be qualitatively extrapolated to other type of superlattices 
where graphene nanoribbons are patterned by selective absorption of different atoms. Another interesting candidate, 
with similar or even better characteristics as graphane is graphene fluoride, a two-dimensional material that has been 
recently synthesized.\cite{cheng:2010, robinson:2010}

\section{Acknowledgments}

This work was supported by the Flemish Science Foundation (FWO-VI), the Belgian Science Policy (IAP), and the
collaborative project FWO-MINCyT (FW/08/01). A. D. H. acknowledges also support from ANPCyT (under Grant No. PICT2008-2236).\\

\appendix

\section{Tight-binding model of graphene nanoribbons}
In order to obtain extra insight in the presented ab initio results for the electronic structure of graphene/graphane nanoribbons, 
we construct a simple tight-binding model. Here we focus only on the graphene part of the nanoribbon. To construct the 
tight-binding Hamiltonian we consider a tight-binding Bloch function per atom in the primitive cell
\begin{eqnarray}
\psi_{\vec{k}}^i (\vec{r})= \frac{1}{\sqrt{N}} \sum_i^N \varphi(\vec{r}-\vec{r}_i)\exp(i\vec{k}\cdot\vec{r}_i),
\end{eqnarray}
where $\vec{r}_i$ is the position of a carbon atom in the primitive cell, and $\varphi$ denotes the $p_z$-like atomic 
wavefunction of a carbon atom. $N$ is the number of considered unit cells. All carbon atoms are positioned on the (unrelaxed) 
honeycomb lattice as in pure graphene. We consider a zigzag nanoribbon with the same number $N_c$ of carbon atoms in the unit 
cell as in the graphene/graphane nanoribbons studied in this paper. For two $\alpha$ edges $N_c=12$ , for a nanoribbon 
with one $\alpha$ and one $\beta$ edge $N_c=13$, and $N_c=14$ in case of two $\beta$ edges. 
The total eigenfunction is then given by
\begin{equation}
\Psi_{\vec{k}}(\vec{r}) = \sum_{i=1}^{N_c} C_i \psi_{\vec{k}}^i (\vec{r}). \label{coefs}
\end{equation}

This results in symmetric $N_c\times N_c$ tight-binding Hamiltonians with the following non-zero off-diagonal elements 
(with only nearest neighbor interactions) (1) nanoribbon with two $\alpha$ edges
\begin{eqnarray}
H(2i+1,2i+2) & =  & \gamma\cos(k c) \;\; \mbox{with } i=0\ldots 5, \\
H(2i+2,2i+3) & = & \gamma\;\; \mbox{with } i=0\ldots 4,
\end{eqnarray}
(2) nanoribbon with one $\alpha$ and one $\beta$ edge
\begin{eqnarray}
H(2i+1,2i+2) & =  & \gamma\cos(k c) \;\; \mbox{with } i=0\ldots 5, \\
H(2i+2,2i+3) & = & \gamma\;\; \mbox{with } i=0\ldots 5,
\end{eqnarray}
(3) nanoribbon with two $\beta$ edges
\begin{eqnarray}
H(2i+1,2i+2) & =  & \gamma \;\; \mbox{with } i=0\ldots 6, \\
H(2i+2,2i+3) & = & \gamma\cos(k c) \;\; \mbox{with } i=0\ldots 5.
\end{eqnarray}

We have chosen the on-site energy as the zero of energy $\int \varphi^*(\vec{r}-\vec{r}_i) H \varphi(\vec{r}-\vec{r}_i) d\vec{r}=0$, 
and the tight-binding parameter $\gamma$ is defined in the usual way
\begin{equation}
\gamma = \int \varphi^*(\vec{r}-\vec{r}_i) H \varphi(\vec{r}-\vec{r}_i - \vec{R}) d\vec{r} \equiv 3 \mbox{ eV}
\end{equation}
where $\vec{R}$ is the vector connecting an atom with its nearest neighbor.

\begin{widetext}
\begin{figure}[h]
\centering
\includegraphics[width=2.0\linewidth]{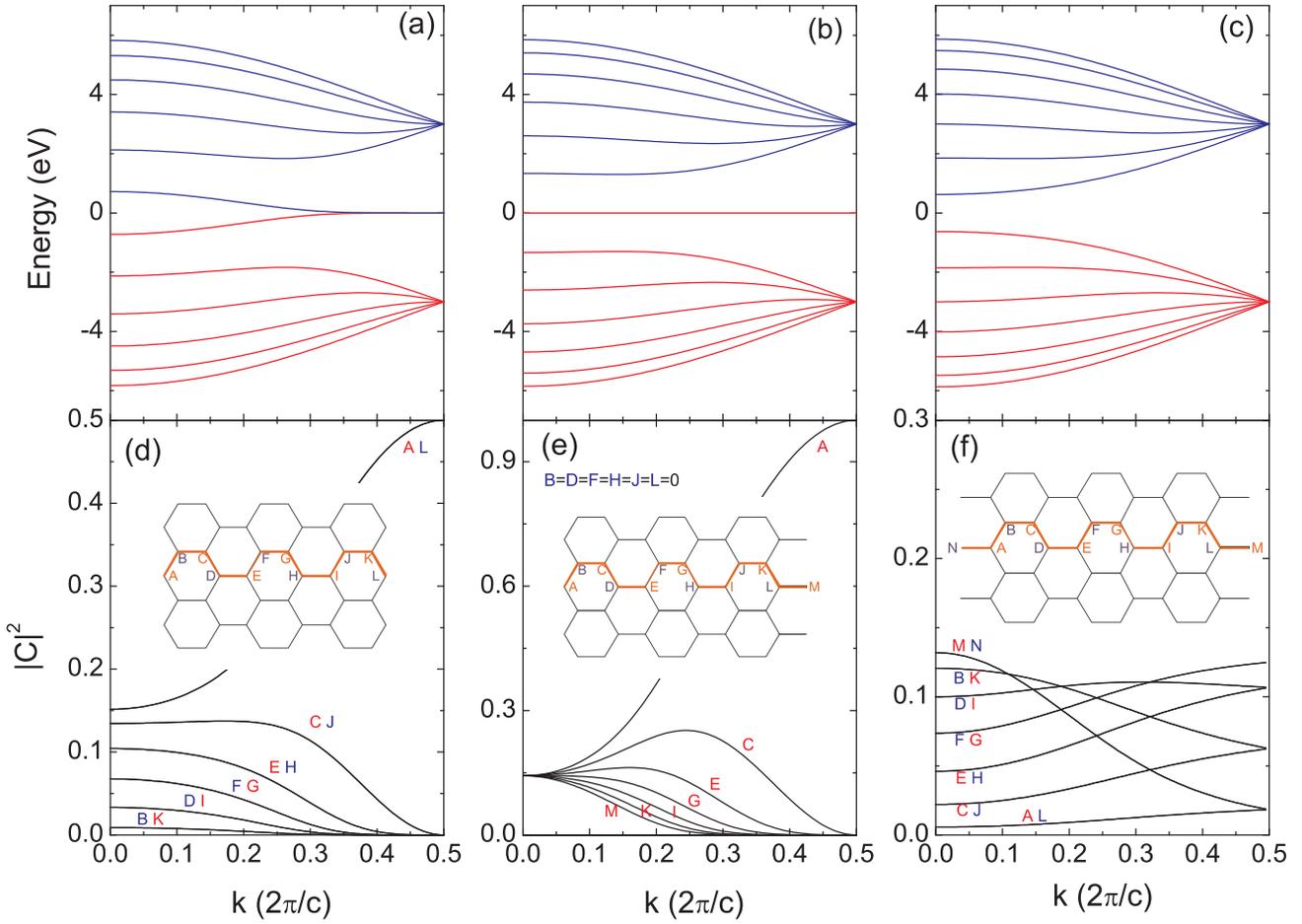}
\caption{(Color online) Band structures for graphene nanoribbons with (a) two $\alpha$ edges containing 12 carbon atoms in the unit cell, 
(b) one $\alpha$ and one $\beta$ edges and 13 atoms in the unit cell, and (c) two $\beta$ edges with 14 atoms in the unit cell. 
The Fermi energy is set to zero. (d), (e), and (f) show the values of the 
coefficients $|C_i|^2$, for all atoms in the unit cell, of the highest valence band corresponding to (a), 
(b), and (c), 
respectively.}
\label{Graph9}
\end{figure}
\end{widetext}

The band structures of the zigzag graphene nanoribbons obtained within this tight-binding model are shown in Fig.~\ref{Graph9}. 
Figures~\ref{Graph9}(a)\--\ref{Graph9}(c) show results for the $\alpha\alpha$, $\alpha\beta$ and 
$\beta\beta$ cases, respectively. The Fermi energy lies at $E_F=0$. The obtained spectra agree qualitatively with the corresponding 
ab initio ones for the non-magnetic graphene/graphane nanoribbons in Figs.~\ref{bands_AA}(c), \ref{bands_AB}(b), and 
\ref{bands_BB}(b). Note the absence of dispersion of the band at $E=0$ for the $\alpha\beta$ graphene nanoribbon 
[Fig.~\ref{Graph9}(b)], in contrast to the ab initio result where a small dispersion was observed, see Fig.~\ref{bands_AB}(d). 
This dispersion is a consequence of the graphane edge, which is not taken into account in the present tight-binding model.

Below each band structure of Fig.~\ref{Graph9}, we plot the corresponding coefficients $|C_i|^2$ from Eq.~(\ref{coefs}) of 
 the highest valence band eigenstate. Each curve is labeled by a letter indicating the corresponding atom in the nanoribbon unit 
cell. Figure~\ref{Graph9}(d) shows that in case of an $\alpha\alpha$ nanoribbon, the highest valence band state is located at 
both edges of the nanoribbon, represented by the carbon atoms A and L. The next most important contribution is not located at 
carbon atoms B and K, but at atoms C and J. Note that these two atoms (C and J) belong to different sublattices, and that C and A 
(and also J and L) belongs to the same sublattice. The results of Fig.~\ref{Graph9}(d) can be correlated with the ab initio results 
of the PCD represented in Figs.~\ref{PCD}(a) and \ref{PCD}(b).

On the other hand, for the case of an $\alpha\beta$ nanoribbon [Fig.~\ref{Graph9}(b)], the dispersionless band at the Fermi level 
is entirely built up from atoms that belong to one sublattice. The contributions coming from the other sublattice given by the 
coefficients B, D, F, H, J, and L are all zero. This is also the case for the ferromagnetic ground state of the 
$\alpha\beta$ nanoribbon as can be seen in the integrated local density of states of Figs.~\ref{PCD}(c) and \ref{PCD}(d).

The case of a nanoribbon with two $\beta$ edges is similar to the $\alpha\alpha$ nanoribbon. The highest valence band is again 
built up from the edge atoms M and N, see Fig.~\ref{Graph9}(f). While the contribution from the neighboring atoms belonging to 
the other sublattice (atoms A and L, or C and J) is again rather small.

The present tight-binding model can be extended to describe the magnetic properties of the system, for example, 
by using the Hubbard model, as it was discussed in Ref.~\onlinecite{kusakabe:2003}.


\begin{thebibliography}{99}

\bibitem{novoselov:2004} K. S. Novoselov, A. K. Geim, S. V. Morozov, D. Jiang, Y. Zhang, S. V. Dubonos, I. V. Grigorieva, 
and A. A. Firsov, Science {\bf 306}, 666 (2004).

\bibitem{han:2007} M. Y. Han, B. \"{O}zyilmaz, Y. Zhang, and P. Kim, Phys. Rev. Lett. {\bf 98}, 206805 (2007).

\bibitem{li:2008} X. Li, X. Wang, L. Zhang, S. Lee, and H. Dai, Science {\bf 319}, 1229 (2008).

\bibitem{han:2010} M. Y. Han, J. C. Brant, and P. Kim, Phys. Rev. Lett. {\bf 104}, 056801 (2010).

\bibitem{jiao:2010} L. Jiao, X. Wang, G. Diankov, H. Wang, and H. Dai, Nature Nanotechnology {\bf 5}, 321 (2010).

\bibitem{trombos:2007} N. Tombros, C. Jozsa, M. Popinciuc, H. T. Jonkman, and B. J. van Wees, Nature (London) {\bf 448}, 571 (2007).

\bibitem{karpan:2007} V. M. Karpan, G. Giovannetti, P. A. Khomyakov, M. Talanana, A. A. Starikov, M. Zwierzycki, 
J. van den Brink, G. Brocks, and P. J. Kelly, Phys. Rev. Lett. {\bf 99}, 176602 (2007).  

\bibitem{son:2006a} Y. W. Son, M. L. Cohen, and S. G. Louie, Phys. Rev. Lett. {\bf 97}, 216803 (2006).

\bibitem{kusakabe:2003} K. Kusakabe and M. Maruyama, Phys. Rev. B {\bf 67}, 092406 (2003).

\bibitem{kan:2008} Er-jun Kan, Z. Li, J. Yang, and J. G. Hou, J. Am. Chem. Soc. 130, 4224 (2008).

\bibitem{son:2006b} Y. W. Son, M. L. Cohen, and S. G. Louie, Nature (London) {\bf 444}, 347 (2006).

\bibitem{koskinen:2009} P. Koskinen, S. Malola, and H. H\"{a}kkinen, Phys. Rev. B {\bf 80}, 073401 (2009).

\bibitem{singh:2009} A. K. Singh and B. I. Yakobson, Nano Letters {\bf 9}, 1540 (2009).

\bibitem{lebegue:2009} S. Leb\`egue, M. Klintenberg, O. Eriksson, and M. I. Katsnelson, Phys. Rev. B {\bf 79}, 245117 (2009).

\bibitem{sofo:2007} J. O. Sofo, A. S. Chaudhari, and G. D. Barber, Phys. Rev. B {\bf 75}, 153401 (2007).

\bibitem{Elias:2009} D. C. Elias, R. R. Nair, T. M. G. Mohiuddin, S. V. Morozov, P. Blake, M. P. Halsall, 
A. C. Ferrari, D. W. Boukhvalov, M. I. Katsnelson, A. K. Geim, and K. S. Novoselov, Science {\bf 323}, 610 (2009).

\bibitem{li:2009} Y. Li, Z. Zhou, P. Shen, and Z. Chen, J. Phys. Chem C {\bf 113}, 15043 (2009).

\bibitem{lu:2009} Y. H. Lu and Y. P. Feng, J. Phys. Chem C {\bf 113}, 20841 (2009).

\bibitem{balog:2010} R. Balog, B. J\o{}rgensen, L. Nilsson, M. Andersen, E. Rienks, M. Bianchi, M. Fanetti, E. L\ae{}gsgaard, 
A. Baraldi, S. Lizzit, et al., Nature Materials {\bf 9}, 315 (2010).

\bibitem{loss:2010} M. J. Schmidt and D. Loss, Phys. Rev. B {\bf 81}, 165439 (2010).

\bibitem{tozzini:2010} V. Tozzini and V. Pellegrini, Phys. Rev. B {\bf 81}, 113404 (2010).

\bibitem{cheng:2010} S. H. Cheng, K. Zou, F. Okino, H. R. Gutierrez, A. Gupta, N. Shen, P. C. Eklund, J. O. Sofo and, J. Zhu, 
Phys. Rev. B {\bf 81}, 205435 (2010).

\bibitem{robinson:2010} J. T. Robinson, J. S. Burgess, C. E. Junkermeier, S. C. Badescu, T. L. Reinecke, F. K. Perkins, 
M. K. Zalalutdniov, J. W. Baldwin, J. C. Culbertson, P. E. Sheehan, and E. S. Snow, Nano Letters {\bf 10}, 3001 (2010).

\bibitem{QE} 
P. Giannozzi, S. Baroni, N. Bonini, M. Calandra, R. Car, C. Cavazzoni, D. 
Ceresoli, G. L. Chiarotti, M. Cococcioni, I. Dabo, A. Dal Corso, S. 
Fabris, et al., J. Phys. Condens. Matter {\bf 21}, 395502 (2009). 
\newblock {\tt (http://www.quantum-espresso.org)} 

\bibitem{pbe} J. P. Perdew, K. Burke, and M. Ernzerhof, Phys. Rev. Lett. {\bf 77}, 3865 (1996).

\bibitem{vanderbilt} D. Vanderbilt, Phys. Rev. B {\bf 41}, 7892 (1990).

\bibitem{jungwirth} T. Jungwirth, Jairo Sinova, J. Ma\u{s}ek, J. Ku\u{c}era, and A. H. MacDonald, Rev. Mod. Phys. {\bf 78}, 809 (2006).

\end{thebibliography}
\end{document}